 \documentstyle[14pt]{article}
\newenvironment%
     {abst}%
     {\normalsize \begin{center}%
     {\bf Abstract.} \end{center}
     \baselineskip 12pt
     \leftmargin 3pc
     \rightmargin 3pc
    \quotation}%
     {\endquotation}%
\newcommand{\be}{\begin{equation}}
\newcommand{\ee}{\end{equation}}
\newcommand{\bea}{\begin{eqnarray}}
\newcommand{\eea}{\end{eqnarray}}
\newcommand{\vac}{\left.\mid 0\right>}
\newcommand{\k}{\vec{k}}
\newcommand{\q}{\vec{q}}
\newcommand{\x}{\vec{x}}
\newcommand{\y}{\vec{y}}
\newcommand{\p}{\vec{P}}

\newcommand{\A}{\tilde{A}}
\newcommand{\B}{\tilde{B}}
\newcommand{\Vac}{\left.\mid 0\right>_{B\B}}
\begin{document}
\thispagestyle{empty}
\begin{flushright}
ISU-IAP.Th95-02, \\
Irkutsk
\end{flushright}

\begin{center}
{\large \bf One-particle excitations and bound states \\
in non-relativistic current $\times$ current model"}.
\footnote{This work is partially supported by RFFR N 94-02-05204
and by grant "Universities of Russia", (St-Pb)
N 94-6.7-2057.}\\
\vspace{.5cm}
{\sl A.N.Vall$^\dagger$, S.E.Korenblit$^*$, V.M.Leviant$^*$,
A.V.Sinitskaya$^\dagger$}. \\
\vspace{1cm}
$\dagger$ { \sl  Department of Theoretical Physics,
                 Irkutsk State University. \\
$*$ \hspace{.4cm} Institute of Applyed Physics,
                  Irkutsk State University, \\
                 Irkutsk, 664003, RUSSIA.}\\
                 e-mail: VALL@physdep.irkutsk.su \\
\end{center}

{\abst

Vacuum structure, one-particle excitations' spectra and bound states
of these excitations are studied in frame of non-relativistic quantum
field model with current $\times$ current type interaction. Hidden symmetry
of the model is found. It could be broken or exact dependign on
the coupling constant value. The effect of "piercing" vacuum , generating
the appearance of heavy fermionic excitations, could occur in the
spontaneously broken phase.

}
\newpage
\centerline{\bf Introduction.}
\vspace{.5cm}

There are several reasons for, - why nonrenormalizable models with
four-fermionic interaction are intensively studied last years \cite{klev}.
First of all, they describe low-energy limit of corresponding gauge theories,
in particular QCD \cite{volkov}, inheriting their global symmetries.
They admit partial bosonization \cite{ebert}, allowing to investigate
meson's spectra, polarizabilities and mesonic scattering lengths.
More over, it turned out they describe good enough "underthreshold"
region i.e., bound states of fermions (e.g., quarks \cite{eb}), and
what is more important, contain a mechanism of forming the physical
vacuum. Nambu, Jona-Lasinio, Vaks and Larkin \cite{njl} were the first
to point out this fact. Using the results of superconductive theory
they described spontaneous breaking of chiral symmetry and found a mechanism
for dynamical generation of masses. The fact of unitary -inequivalent
representations of canonical anticommutation relations \cite{h,bsh,grib}
had been essentially used by them. This fact, in its turn, opened
the possibility to study vacuum structure in the theories with four -
fermionic interaction both in the usual Minkovsky space \cite{ali}
and on lattice \cite{khan}

Note, that in the relativistically invariant theories any quantum
field contains creation and annihilation operators, and Hamiltonian
always includes so-called "fluctuation" terms \cite{schveb}.
This kind of operators generate unitary - inequivalent transformations
relevant to the reconstruction of basic state of quantum field theory -
vacuum. Renormalization group transformations, belonging to the same unitary -
inequivalent type \cite{efimov}, serve in a sense as their "compensators".

The presence of such terms in Hamiltonian results to one more difficulty
when describing bound states: it is impossible to write down self-consistent
Bethe-Salpeter equation. Namely by this reason the conventional approach
to describing bound states is based on the solution of the sets of
Bethe-Salpeter and Schwinger-Dyson equations, because the first set
contains total two - point Green functions. By the same reason it is
impossible in Haag expansion \cite{haag,green,um} for heisenberg field
over degrees of normal ordered "in" (or "out") operators,
to write down the equations on coefficient functions, because there always
would be contributions from high operator monomials to low ones \cite{green2}.

In this paper we use non-relativistic model of interacting "singlet"
fermions (fermions with a single helicity)
possessing the isotopic spin, to investigate the structure of:
vacuum, one-particle and two-particle states. This model is a non -
relativistic limit of the "B" model \cite{domitr} that has a number
of interesting peculiarities. In particular, constant contributions
to one-particle spectra of particle and antiparticle differ by
sign \cite{our}. As we will show later, it is related to the answer -
in what phase (broken or not) the model is considered.

We will formulate the conditions, when the "fluctuation" part is absent,
which allow to write the equations on bound states and to solve them
analytically.
\vspace{.5cm}

\centerline{\bf 2. Choice of model.}
\vspace{.5cm}

Consider Hamiltonian with current $\times$ current interaction
      \be
 H = \int d^3x \left[\Psi^{\dagger}_\alpha (x)\varepsilon (\hat{\vec p})
    \Psi_\alpha(x)
   - \lambda J^\mu(x)J_\mu(x)\right] , \label{h}
       \ee
where
\bea
J^0 &=& \Psi^\dagger_\alpha(x)\Psi_\alpha(x) \nonumber \\
\vec{J}(x) &=& \frac{1}{2mc}\left(\Psi^\dagger_\alpha(x)\hat{\vec
p}\Psi_\alpha(x) -
\hat{\vec p}\Psi^\dagger_\alpha(x) \Psi_\alpha(x)\right),\;\;\hat{\vec p} =
-i\vec{\nabla}, \nonumber \\
\varepsilon (\hat{\vec p})e^{i\k\x} &=& \varepsilon (\k)e^{i\k\x}, \nonumber
\eea
$\Psi_\alpha (x)$ is a fermionic field,
 $\alpha = 1,2$ is isospin index,
$\varepsilon (\k) = \frac{\k^2}{2m} + mc^2$ - "bare" spectrum of free
fermions.
As simple analysis shows, there exist two independent realizations of
the heisenberg field $\Psi(\x, 0)$
\bea
\Psi_1(\x, 0) &=&
\frac{1}{(2\pi)^{\frac{3}{2}}} \int d^3k f(\k) e^{i\k\x}
 A_\alpha (\k) \nonumber \\
\Psi_2(\x, 0) &=&
\frac{1}{(2\pi)^{\frac{3}{2}}} \int d^3k g(\k) e^{-i\k\x}
 \tilde{A}_\alpha^\dagger (\k) \label{sols}
\eea
such that the vacuum states $\vac_A$ and $\vac_{\tilde{A}}$
and corresponding one-particle excitations
$A^\dagger_\alpha\vac_A , \tilde{A}^\dagger_\alpha
\vac_{\tilde{A}}$ are the eigenstates of Hamiltonian (\ref{h}).
One can check, the states $\vac_A$ and $\vac_{\tilde{A}}$ are mutually
orthogonal
and the energy spectra of excitations $A$ and $\tilde{A}$ are different. We
can formally unite these solutions (\ref{sols}) into one field
$\Psi(\x, 0)$:
\be
\Psi(\x, 0) \propto
\frac{1}{(2\pi)^{\frac{3}{2}}} \int d^3k
\left[
\left(
\begin{array}{c}
  f(\k) \\
   0
\end{array}\right)e^{i\k\x} A_\alpha (\k) +
\left(
\begin{array}{c}
   0 \\
   g(\k)
\end{array}\right)e^{-i\k\x} \tilde{A}^\dagger_\alpha (\k) \right].
\label{psi}
\ee
After replacement $
\left(\begin{array}{c}
  f(\k) \\
   0
\end{array}  \right) \rightarrow f^a(\k),
\left(
\begin{array}{c}
   0 \\
   g(\k)
\end{array}\right) \rightarrow g^a(\k) $ field $\Psi_\alpha (\x, 0)$
acquires index  $"a"$; let it run  the values $1,2,\dots $.

Thus, we want to combine these two sets of operators, corresponding to the
two independent solutions, demanding them to be described by a single
field $\Psi^a_\alpha (\x,0)$. The most simple generalization of the
Hamiltonian (\ref{h}) can be obtained from the following Lagrangian
density
\be
{\cal L}(x) = \Psi^{\dagger a}_\alpha (x)\left(i\frac{\partial}{\partial t} -
          \varepsilon (\hat{\vec p}) \right) \Psi^a_\alpha (x) + \lambda
J^{\mu}(x) J_{\mu} (x)
\label{L}
\ee
where
   $$
J^0(x) = \Psi^{+a}_\alpha (x)\Psi^a_\alpha(x)
   $$
   $$
         \vec{J} = \frac{1}{2mc}\left(\Psi^{+a}_\alpha (x)\hat{\vec p}\cdot
                   \Psi^a_\alpha(x)
             - \hat{\vec p}\Psi^{+a}_\alpha (x)\cdot\Psi^a_\alpha(x) \right)
   $$
 From here by standard way we get the expression for Hamiltonian
      \be
H = \int d^3x \left[\Psi^{+a}_\alpha (x)\epsilon(\hat{\vec p})\Psi^a_\alpha(x)
   - \lambda J^\mu(x)J_\mu(x)\right] \equiv H_N + H_{Fl}\label{H}
       \ee
and heisenberg equation corresponding to this Hamiltonian
\bea
i\frac{\partial}{\partial t}\Psi^a_\alpha (x) &=& [\Psi^a_\alpha (x), H] =
\label{he} \\
=\varepsilon (\hat{\vec p})\Psi^a_\alpha (x) - \lambda \left\{
\Psi^a_\alpha (x), j^0\right\}
&-&\frac{\lambda}{mc}\biggl\{\bigl(\hat{\vec p}
\Psi^a_\alpha (x)\bigr), \vec{J}\biggr\} - \frac{\lambda}{2mc}
\left\{\Psi^a_\alpha (x), \hat{\vec p}\vec{J}\right\} \nonumber
\eea

Let us take the Fock representation of the heisenberg field
$\Psi^a_\alpha (x)$
at $t =0$ as an expansion over creation and annihilation operators
\be
\Psi^a_\alpha (\x,0) = \frac{1}{(2\pi)^{\frac{3}{2}}}\int d^3k \left( f^a(\k)
e^{i\k\x}A_\alpha (\k) + g^a(\k)e^{-i\k\x}\tilde{A}^\dagger_\alpha
(\k)\right), \;\;\;a,b=1,2,...\label{rep}
\ee
with $\left\{\tilde{A}_\alpha (\k), \tilde{A}^\dagger_\beta (\q)\right\} =
\left\{A_\alpha (\k), A^\dagger_\beta (\q)\right\} = \delta_{\alpha\beta}
\delta (\k - \q)$, all other anticommutators are equal to zero.
The canonical quantization of field $\Psi^a_\alpha (\x,0)$ with
Lagrangian (\ref{L}) requires  carrying out of relations
\be
\left\{\Psi^a_\alpha (\x,t_x), \Psi^b_\beta (\y,t_y)\right\}\delta(t_x - t_y) =
\delta_{\alpha\beta} \delta_{ab} \delta^4(x - y).
\label{car}
\ee
Inserting here the representation (\ref{rep}) we obtain  "canonical"
constrains for the amplitudes $f^a (\k)$ and $g^a (\k)$:
\be
f^a (\k)\bar{f}^b (\k) + g^a (-\k)\bar{g}^b (-\k) = {\bf I}^{ab}, \;\;\;
a,b = 1,2,...\label{con}
\ee
${\bf I}$ is unit matrix. The constrains (\ref{con}), which can be
interpreted as a "decomposition of unit" over the amplitudes
$f^a (\k)$ and $g^a (\k)$, impose on the latter certain restrictions.
Remarkably, nevertheless, that the relations (\ref{con})  strictly fix
permissible dimensions of degrees of freedom i.e., the value of $Sp{\bf I}$.
Indeed, from (\ref{con}) it is easy to obtain:
\bea
f^a (\k)\left[1 - \sum_{b}\mid f^b (\k)\mid^2\right] &=& g^a(-\k)
\sum_{b}f^b (\k)\bar{g}^b(-\k) \nonumber \\
g^a(\k)\left[1 - \sum_{b}\mid g^b(\k)\mid^2\right] &=& f^a (-\k)
\sum_{b}\bar{f}^b (-\k)g^b(\k) \label{ab} \\
\sum_{a}\mid f^a (\k)\mid^2 + \sum_{a}\mid g^a (-\k)\mid^2 &=& Sp{\bf I}
\nonumber
\eea
Let $\sum_{b}f^b (\k)\bar{g}^b(-\k) \not= 0$, then from (\ref{ab}) follows
$Sp{\bf I} = 1$, i.e. $a,b = 1$.

Let now
\be
\sum_{b}f^b (\k)\bar{g}^b(-\k) = 0, \label{ort}
\ee
then we see that  $\sum_{a}\mid f^a (\k)\mid^2 =
\sum_{a}\mid g^a (\k)\mid^2 =1$ and, hence
$Sp{\bf I} = 2$ i.e., $a,b = 1,2$. So, the realization of the canonical
relations (\ref{car}) in the representation
(\ref{rep}) is possible only in one - or two - dimensional spaces of
amplitudes $f^a (\k)$ and $g^a (\k)$, moreover in the last case
it is necessary to perform the orthogonality condition (\ref{ort}).
Further on we will consider two-component model, because $Sp{\bf I} = 1$
case is reduced to the solution (\ref{sols}).

Let us obtain the expression of the Hamiltonian in Fock representation.
To that end we insert decomposition (\ref{rep}) into (\ref{H}), then
\[
H = H_N + H_{Fl}
\]
\vspace{-.5cm}
\bea
H_N &=& \int d^3k\left[\epsilon(\k)A^+_\alpha(\k)A_\alpha(\k)
       +\epsilon(-\k)\A_\alpha(\k)\A^+_\alpha(\k)\right] - \nonumber \\
    &-& \lambda \frac{1}{(2\pi)^3}\int d^3k_1d^3k_2d^3k_3d^3k_4
          \biggl\{\biggr. \delta(\k_1 - \k_2 + \k_3 - \k_4)\nonumber \\
   &\;&  \bar{f}^a(\k_1)f^a(\k_2)\bar{f}^b(\k_3)f^b(\k_4)
          \left(1 - \frac{(\k_1+\k_2)(\k_3+\k_4)}{4m^2c^2}\right)\times
        \nonumber \\
   &\times& \left[-A_\alpha^+(\k_1)A_\beta^+(\k_3)A_\alpha(\k_2)A_\beta(\k_4)
       +  \delta(\k_2 - \k_3)A_\alpha^+(\k_1)A_\alpha(\k_4)\right] +
           \nonumber \\
   &+&   \delta(\k_1 - \k_2 + \k_3 - \k_4)
          \bar{g}^a(\k_1)g^a(\k_2)\bar{g}^b(\k_3)g^b(\k_4)
          \left(1 - \frac{(\k_1+\k_2)(\k_3+\k_4)}{4m^2c^2}\right)\times
           \nonumber \\
   &\times& \left[-\A_\alpha^+(\k_2)\A_\beta^+(\k_4)\A_\alpha(\k_1)
            \A_\beta(\k_3) +
          \delta(\k_1 - \k_4)\A_\alpha^+(\k_2)\A_\alpha(\k_3)\right. -
         \nonumber \\
    &-&   2\delta(\k_3 - \k_4)\A_\alpha^+(\k_2)\A_\alpha(\k_1) -
         2\delta(\k_1 - \k_2)\A_\alpha^+(\k_4)\A_\alpha(\k_3) +\nonumber \\
      &+& \left. 4\delta(\k_1 - \k_2)\delta(\k_3 - \k_4) \right] + \nonumber \\
    &+& 2\delta(\k_3 - \k_4 - \k_1 + \k_2)
        \bar{f}^a(\k_1)f^a(\k_2)\bar{g}^b(\k_3)g^b(\k_4)
        \left(1 + \frac{(\k_1+\k_2)(\k_3+\k_4)}{4m^2c^2}\right)\times
        \nonumber \\
   &\times& \left[A_\alpha^+(\k_1)\A_\beta^+(\k_4)A_\alpha(\k_2)\A_\beta(\k_3)
       +  2\delta(\k_3 - \k_4)A_\alpha^+(\k_1)A_\alpha(\k_2)\right] +
           \nonumber \\
   &+&   \delta(\k_1 + \k_2 - \k_3 - \k_4)
          \bar{f}^a(\k_1)g^a(\k_2)\bar{g}^b(\k_3)f^b(\k_4)
          \left(1 + \frac{(\k_1-\k_2)(\k_3-\k_4)}{4m^2c^2}\right)\times
           \nonumber \\
   &\times& A_\alpha^+(\k_1)\A_\alpha^+(\k_2)\A_\beta(\k_3)
             A_\beta(\k_4) +  \nonumber\\
   &+&   \delta(\k_1 + \k_2 - \k_3 - \k_4)
          \bar{f}^a(\k_1)g^a(\k_2)\bar{g}^b(\k_3)f^b(\k_4)
          \left(1 + \frac{(\k_1-\k_2)(\k_3-\k_4)}{4m^2c^2}\right)\times
           \nonumber \\
   &\times& \left[A_\alpha^+(\k_1)\A_\alpha^+(\k_2)\A_\beta(\k_3)
             A_\beta(\k_4) -
          \delta(\k_2 - \k_3)A_\alpha^+(\k_1)A_\alpha(\k_4)\right. -
         \nonumber \\
    &-&  \left.\left.\delta(\k_1 - \k_4)\A_\alpha^+(\k_2)\A_\alpha(\k_3) +
         2\delta(\k_1 - \k_4)\delta(\k_2 - \k_3) \right]\right\}
\label{HN}
    \eea
    \bea
    H_{Fl} &=& - \lambda \frac{1}{(2\pi)^3}\int d^3k_1d^3k_2d^3k_3d^3k_4
          \biggl\{\biggr. \delta(\k_1 - \k_2 + \k_3 + \k_4)\nonumber \\
   &\;&   \bar{f}^a(\k_1)f^a(\k_2)\bar{f}^b(\k_3)g^b(\k_4)
          \left(1 - \frac{(\k_1+\k_2)(\k_3-\k_4)}{4m^2c^2}\right)\times
        \nonumber \\
   &\times&
\left[2A_\alpha^+(\k_1)A_\beta^+(\k_3)\A_\beta^+(\k_4)A_\alpha(\k_2)
       +  \delta(\k_2 - \k_3)A_\alpha^+(\k_1)\A_\alpha^+(\k_4)\right] +
           \nonumber \\
   &+&   \delta(\k_1 + \k_2 - \k_3 + \k_4)
          \bar{g}^a(\k_1)f^a(\k_2)\bar{f}^b(\k_3)f^b(\k_4)
          \left(1 - \frac{(\k_2-\k_1)(\k_3+\k_4)}{4m^2c^2}\right)\times
           \nonumber \\
   &\times& \left[2\A_\beta^+(\k_3)\A_\alpha(\k_1)A_\alpha(\k_2)
            A_\beta(\k_4) +
          \delta(\k_2 - \k_3)\A_\alpha(\k_1)A_\alpha(\k_4)\right] -
         \nonumber \\
    &+&  \delta(\k_1 - \k_2 - \k_3 - \k_4)
        \bar{g}^a(\k_1)g^a(\k_2)\bar{f}^b(\k_3)g^b(\k_4)
        \left(1 - \frac{(\k_1+\k_2)(\k_3-\k_4)}{4m^2c^2}\right)\times
        \nonumber \\
   &\times& \left[-2\A_\alpha^+(\k_2)A_\beta^+(\k_3)\A_\beta^+(\k_4)
                 \A_\alpha(\k_1)
       +  \delta(\k_1 - \k_4)\A_\alpha^+(\k_2)A_\alpha^+(\k_3) + \right.
       \nonumber \\
    &+&  \left. 4\delta(\k_1 - \k_2)A_\alpha^+(\k_3)\A_\alpha^+(\k_4)
                    \right] +
           \nonumber \\
   &+&   \delta(\k_1 + \k_2 + \k_3 - \k_4)
          \bar{g}^a(\k_1)f^a(\k_2)\bar{g}^b(\k_3)g^b(\k_4)
          \left(1 + \frac{(\k_1-\k_2)(\k_3+\k_4)}{4m^2c^2}\right)\times
           \nonumber \\
   &\times&
    \left[-2\A_\beta^+(\k_4)\A_\alpha(\k_1)A_\alpha(\k_2)
             \A_\beta(\k_3) +
           \delta(\k_1 - \k_4)A_\alpha(\k_2)\A_\alpha(\k_3) +\right.
           \nonumber \\
   &+&    \left. 4\delta(\k_3 - \k_4)\A_\alpha(\k_1)A_\alpha(\k_2)
          \right] + \nonumber \\
   &+&   \delta(\k_1 + \k_2 + \k_3 + \k_4)
          \bar{f}^a(\k_1)g^a(\k_2)\bar{f}^b(\k_3)g^b(\k_4)
          \left(1 - \frac{(\k_1-\k_2)(\k_3-\k_4)}{4m^2c^2}\right)\times
           \nonumber \\
   &\times&
\left[A_\alpha^+(\k_1)\A_\alpha^+(\k_2)A_\beta^+(\k_3)
             \A_\beta^+(\k_4) \right] \nonumber  \\
   &+&   \delta(\k_1 + \k_2 + \k_3 + \k_4)
          \bar{g}^a(\k_1)f^a(\k_2)\bar{g}^b(\k_3)f^b(\k_4)
          \left(1 - \frac{(\k_1-\k_2)(\k_3-\k_4)}{4m^2c^2}\right)\times
           \nonumber \\
   &\times&
\left.\left[\A_\alpha(\k_1)A_\alpha(\k_2)\A_\beta(\k_3)
             A_\beta(\k_4) \right]\right\}
 \label{HFL}
    \eea

The components where the number of creation operators  is not equal to
the number of annihilation ones are singled out to the "fluctuation part
of the Hamiltonian $H_{Fl}$

It is worth to note here that as a straight consequence of the canonical
relations (\ref{car}), the "fluctuation" part of the kinetic term vanishes
identically. It is always true for space integral from any bi-local field's
form. The same fact and for the same reason takes place in a relativistic
quantum theory. This is not the case when $Sp{\bf I} = 1$.

Define vacuum as a state $\vac_{A\tilde{A}}$ without particles
$A$ and $\tilde{A}$:
\be
A_\alpha\vac_{A\tilde{A}} = \tilde{A}_\alpha\vac_{A\tilde{A}} =  0
\label{vac}
\ee
The presence of "fluctuation" part in the Hamiltonian leads to the fact that
vacuum$\vac_{A\tilde{A}}$ and one-particle excitations $A^\dagger_\alpha(\k)
\vac_{A\tilde{A}}, \tilde{A}^\dagger (\k)\vac_{A\tilde{A}}$ cease to be
the eigenstates of the Hamiltonian. Indeed, action of $H$ to the vacuum
results the state:
\be
H\vac_{A\tilde{A}} = W_0\vac_{A\tilde{A}} + \Delta H(2)\vac_{A\tilde{A}} +
\Delta H(4)\vac_{A\tilde{A}}, \label{Va}
\ee
where two last terms correspond to two- and four- particle states.
Thus, the time evolution of the vacuum state is developed on a background
of production of infinite number of pairs $A\tilde{A}$.
There is one more aspect related to the presence of "fluctuation" terms in
a Hamiltonian. The point is that in this case evolution operator
will contain terms relevant to unitary - inequivalent transformations.
Thus, alteration of transformation parameter (that is time) is accompanied
by the motion over a continuum of orthogonal Hilbert spaces \cite{wight}
and, in general,
is accompanied by continuous dynamical reconstruction of vacuum. Such a
situation, clearly, can not be included in the frameworks of natural idea
about evolution as a development of a system with time in single Hilbert
space.

It is not hard to check that the "fluctuation" terms in $H$ disappear on
the solutions (\ref{con}) when  $f^a (\k), g^a (\k)$ do not depend on
momentum. In this case we have:
\be
H_{Fl}\equiv 0, \;\;\;\;H = H_N \label{Hn}
\ee
Now the vacuum and one-particle excitations
$A^\dagger_\alpha(\k)\vac_{A\tilde{A}}$ and
$\tilde{A}^\dagger (\k)\vac_{A\tilde{A}}$
become eigenstates of $H$, enabling to make further analysis.
It is interesting to note that the solution with constant
$f^a $ and $g^a $ can be achieved by another way, supposing the
 vacuum and one-particle excitations to be determined by the normal
part of the Hamiltonian $H_N$. Then, in the stationary Schredinger equation
$\left[ H_N , A^\dagger_\alpha (\k)\right]\vac =
E_A(\k)A^\dagger_\alpha (\k)\vac $ spectrum $E_A(\k)$ is calculated to be
function of the amplitudes $f^a (\k)$ and $g^a (\k)$. If now we imply
that the "bare" spectrum "dressing" would not change the functional form of
the spectrum, but would provide the "dressing" of mass and energy "gap",
then it could be possible when $f^a, \;\;g^a = const$.
\vspace{.5cm}

\centerline{\bf 3. One-particle excitation's spectra.}
\vspace{.5cm}

Let us now introduce notations we will use in the future.
Since Lagrangian (\ref{L}) is known to be nonrenormalizable it is
necessary to make use of an ultraviolet cut-off $\Lambda$.
According to this we introduce the notations:
\be
\frac{1}{(2\pi)^3}\int \limits^\Lambda d^3k \equiv \frac{1}{V^*},\;\;\;
<k^2> \equiv \frac{\int \limits^\Lambda \k^2 d^3k}{\int\limits^\Lambda d^3k},
\;\;\;g \equiv \frac{\lambda}{V^*} \label{def}
\ee
Physical meaning of $V^*,\mbox{and}\;<k^2>$ quantities may be traced by
inputting into the representation (\ref{rep}) or into the interaction
(\ref{L}) a formfactor. Then, it is easy to verify:
$V^*$ is space volume of one-particle excitation, and
$<k^2>$ is the average momentum within this volume. As will be shown below
these quantities are actually determined by Compton length of respective
excitation mass of "dressed" fermion.  Renormalized coupling constant
$g$ has dimension of energy and enters alone into the final expressions
for the all dynamical characteristics.

Performing integration in relations (\ref{def}) we obtain:
\be
\Lambda^2 = \frac{5}{3}<k^2>,\;\;\;V^{*-1} = \frac{1}{6\pi^2}\Lambda^3,\;\;\;
\lambda\Lambda^3 = 6\pi^2g.      \label{gef2}
\ee

In order to find one-particle spectrum consider stationary Schredinger
equation
\bea
\left[ H , A^\dagger_\alpha (\k)\right]\vac &=&
E_A(\k)A^\dagger_\alpha (\k)\vac \nonumber \\
\left[ H , \tilde{A}^\dagger_\alpha (\k)\right]\vac &=&
E_{\tilde{A}}(\k)\tilde{A}^\dagger_\alpha (\k)\vac \label{E1}
\eea
Now inserting Hamiltonian (\ref{H}) into (\ref{E1}) on the conditions
(\ref{Hn}) we get:
\bea
E_A(\k) &=& \varepsilon(\k) + \frac{g}{4m^2c^2}k^2 -5g + g\dot\frac{
            <k^2>}{4m^2c^2} \nonumber \\
E_{\tilde{A}}(\k) &=& -\varepsilon(\k) + \frac{g}{4m^2c^2}k^2 +3g +
                  g\dot\frac{<k^2>}{4m^2c^2} \label{E2} \\
\varepsilon(\k) &=& \frac{k^2}{2m} + mc^2 \nonumber
\eea
 From the equation $H\vac_{A\tilde{A}} = W_0\vac_{A\tilde{A}}$ for the
energy density  of vacuum there follows:
\be
\frac{V^*}{V}W_0 = 2<\varepsilon (\k)> - 4g = \frac{<k^2>}{m} + 2mc^2 -
                  4g, \label{W}
\ee
and for $E_A(\k)$:
\bea
E_A(\k) &=& \frac{k^2}{2m_A} + E_A(0), \;\;\;E_A(0) = mc^2 -5g
          + g\frac{<k^2>}{4m^2c^2} \label{E3}\\
m_A &=& \frac{m}{1 + \frac{g}{2mc^2}} \nonumber
\eea
$E_A(0)$ determines energy "gap".  At the absence of interaction the
energy "gap" coincides  with the rest-frame energy $mc^2$.
However, when the interaction is "switched-on" $E_A(0)\not= m_Ac^2$
generally speaking. The equality is possible in a case we will consider
below. Thus, the interaction leads to renormalization of mass and
energy "gap". One can derive expression for the "bare" mass via
"physical" $m_A$ from (\ref{E3})
\be
m = \frac{m_A}{2}\left(1 + \sqrt{1 + \frac{2g}{m_Ac^2}}\right) \label{m}
\ee
We will use this expression to exclude the "bare" mass.

Consider now the spectrum $E_{\tilde{A}}(\k)$. From (\ref{E2}) we have:
\bea
E_{\tilde{A}}(\k) &=& \frac{k^2}{2m_{\tilde{A}}} + E_{\tilde{A}}(0), \;\;\;
E_{\tilde{A}}(0) = - mc^2 + 3g
          + g\frac{<k^2>}{4m^2c^2} \label{E4}\\
m_{\tilde{A}} &=& \frac{m}{\frac{g}{2mc^2} - 1} \nonumber
\eea
 From here follows that an interpretation of exitations $\tilde{A}$
is different depending on the value $\frac{g}{2mc^2}$.
First af all the excitation $\tilde{A}$ can be interpreted as a "hole"
relatively to $A$ -particle only at $g = 0$. In region
$0 < \frac{g}{2mc^2}<1$ the excitation corresponds to "bubble" in
vacuum, for the group velocity and momentum are arrowed to opposite
directions. At  $\frac{g}{2mc^2} = 1$ there takes place a phenomen
called "piercing" of vacuum; and at $\frac{g}{2mc^2} > 1$ the
excitation $\tilde{A}$ becomes real particle with the mass (\ref{E4}).
The expression of "bare" mass via "physical" $m_{\tilde{A}}$ reads:
\be
m = \frac{m_{\tilde{A}}}{2}\left(\sqrt{1 + \frac{2g}{m_{\tilde{A}}c^2}}
- 1   \right) \label{m1}
\ee
Equating $m$ from (\ref{m}) and (\ref{m1}) we find out the relation between
$m_A$ and  $m_{\tilde{A}}$:
\be
m_{\tilde{A}} = \left(1 +
\frac{4}{\alpha}\right)m_A ,\;\;\; \mbox{where} \;\;\;\alpha = \sqrt{1 +
\frac{2g}{m_Ac^2}} - 3. \label{mm}
\ee
The excitation $\tilde{A}$ corresponds to real particle when $\alpha > 0$.
 From the relation (\ref{mm}) it follows that particle $\tilde{A}$
is heavier then particle $A$ at any $\alpha > 0$. Moreover, at
sufficiently small $\alpha$ the $m_{\tilde{A}}$ can be as large as
possible.

One may find the relation between the energy "gaps" using (\ref{E3})
and (\ref{E4}):
\be E_A(0) + E_{\tilde{A}}(0) = 2g +
g\frac{<k^2>}{2m^2c^2}  \label{EE}
\ee
We postpone for the moment further analysis of one-particle excitation's
spectra, but as will be shown below difference of masses
 $\triangle m = m_{\tilde{A}} - m_A = \frac{4}{\alpha}m_A$
is caused by spontaneouse breaking of $SU(2)$  symmetry.

Taking into account the obtained spectra we rewrite Hamiltonian (\ref{H})
in the following compact form:
\bea
H &=& \int d^3k \left[E_A(\k)A^\dagger_\alpha (\k)A_\alpha (\k) +
   E_{\tilde{A}}(\k)\tilde{A}^\dagger_\alpha (\k)\tilde{A}_\alpha (\k)\right] +
\nonumber \\
&+& \lambda \frac{1}{(2\pi)^3}\int d^3k_1d^3k_2d^3k_3d^3k_4 \left\{
\delta(\k_1 - \k_2 + \k_3 - \k_4)\left[1 -
\frac{(\k_1 + \k_2)(\k_3 + \k_4)}{4m^2c^2}\right]\right.  \nonumber   \\
&\times& \left(
A^\dagger_\alpha (\k_1)A^\dagger_\beta (\k_3)A_\alpha (\k_2)A_\beta (\k_4) +
\tilde{A}^\dagger_\alpha (\k_1)\tilde{A}^\dagger_\beta (\k_3)
\tilde{A}_\alpha (\k_2)\tilde{A}_\beta (\k_4) \right. - \nonumber \\
&-&
2\left.\left. A^\dagger_\alpha (\k_1)\tilde{A}^\dagger_\beta (-\k_3)
A_\alpha (\k_2)\tilde{A}_\beta (-\k_4) \right)\right\} + W_0 \label{HH}
\eea
where $E_{A},\; E_{\tilde{A}}$ and $W_0$ are defined above by (\ref{E2})
and (\ref{W}).
\vspace{.5cm}

\centerline{\bf 4. Bound states.}
\vspace{.5cm}

Linear shell of any $n$ - particle Fock column, as can be proved from
representation (\ref{HH}), forms irreducible space of the Hamiltonian
$H$. This fact enables to construct $n$ - particles eigenstates and for
$n = 2$ we have:
\bea
H A^\dagger_\alpha (\q_2)A^\dagger_\beta (\q_1)\vac &=& \left(W_0 +
E_A(\q_1) + E_A(\q_2)\right)A^\dagger_\alpha (\q_2)A^\dagger_\beta (\q_1)\vac
- \nonumber \\
- \lambda \frac{2}{(2\pi)^3}\int d^3k_1d^3k_2
\delta (\k_1 &+& \k_2 - \q_1 - \q_2)
\left[1 - \frac{(\q_1 + \k_1)(\q_2 + \k_2)}{4m^2c^2}\right]\times \nonumber \\
&\times& A^\dagger_\alpha (\k_2)A^\dagger_\beta (\k_1)\vac \nonumber \\
H A^\dagger_\alpha (\q_2)\tilde{A}^\dagger_\beta (\q_1)\vac &=& \left(W_0 +
E_{\tilde{A}}(\q_1) + E_{\tilde{A}}(\q_2)\right)
A^\dagger_\alpha (\q_2)\tilde{A}^\dagger_\beta (\q_1)\vac
+ \nonumber \\
+ \lambda \frac{2}{(2\pi)^3}\int d^3k_1d^3k_2
   \delta (\k_1 &+& \k_2 - \q_1 - \q_2)
\left[1 + \frac{(\q_1 + \k_1)(\q_2 + \k_2)}{4m^2c^2}\right]\times \nonumber \\
A^\dagger_\alpha (\k_2)\tilde{A}^\dagger_\beta (\k_1)\vac \label{two}
\eea
Action of $H$ to the state  $\tilde{A}^\dagger_\alpha\tilde{A}^\dagger_\beta
\vac$ gives the same relust as its action to the state
$A^\dagger_\alpha A^\dagger_\beta \vac$. The presence of
$\delta$ - function in the l.h.s. of (\ref{two}) indicates that
irreducible state is realized on hypersurface
$\q_1 + \q_2 = \vec{P} =
\mbox{const.}$ Therefore, wave function
$D_{\alpha\beta}(\q_1, \q_2)$ of two-particle eigenstate satisfies the
equation:
\bea
H \mid A,A> &=& (W_0 + \mu_A(\vec{P}))\mid A,A>, \;\;\;\mbox{where}
\nonumber \\
\mid A,A> &=& \int d^3q_1 d^3q_2 \delta (\q_1 + \q_2 - \vec{P})
D_{\alpha\beta}(\q_1, \q_2)A^\dagger_\alpha
(\q_2)A^\dagger_\beta (\q_1)\vac .
\label{bs}
\eea
Analogous equations are held for the states
$\mid \tilde{A},\tilde{A}>$ and $\mid A,\tilde{A}>$ with respective
$\mu (\p)$ and $D_{\alpha\beta}(\q_1, \q_2)$.

 From (\ref{two})
one can see that equations on $D_{\alpha\beta}(\q_1, \q_2)$ and $\mu (\p)$
for the states $\mid A, A>$ and $\mid A,\tilde{A}>$ difer by sign of
contribution from the time - component of current $J^0 (x)$ in the
Hamiltonian (\ref{HH}).

Combining (\ref{two}) and (\ref{bs}) we obtain the set of equations:
\bea
D^{AA}_{\alpha\beta}(\q_1, \q_2) &=& \frac{2
F^{AA}_{\alpha\beta}(\q_1, \q_2)}
{E_A(\q_1) + E_A(\q_2) - \mu (\p)} \label{DA} \\
F^{AA}_{\alpha\beta}(\q_1, \q_2) &=&
\lambda \frac{1}{(2\pi)^3}\int d^3k_1d^3k_2
\delta (\k_1 + \k_2 - \p)\times\nonumber \\
&\times& D^{AA}_{\alpha\beta}(\k_1, \k_2)
\left[1 - \frac{(\q_1 + \k_1)(\q_2 + \k_2)}{4m^2c^2}\right]. \nonumber
\eea
Equations on the state $\mid \tilde{A} \tilde{A}>$ looks equally
after substitution $E_A(\q) \rightarrow E_{\tilde{A}}(\q)$ in the propagator.
For the state $\mid A \tilde{A}>$ the equations are derived by
replacement $E_A(\q_1) \rightarrow E_{\tilde{A}}(\q_1)$ in the
expression for $D_{\alpha\beta}(\q_1, \q_2)$ and besides that the
formfactor $F_{\alpha\beta}(\q_1, \q_2)$ has to be changed:
\bea
F^{A\tilde{A}}_{\alpha\beta}(\q_1, \q_2) &=& -
\lambda \frac{1}{(2\pi)^3}\int d^3k_1d^3k_2
\delta (\k_1 + \k_2 - \p) \times \nonumber \\
&\times& D^{A\tilde{A}}_{\alpha\beta}(\k_1, \k_2)
\left[1 + \frac{(\q_1 + \k_1)(\q_2 + \k_2)}{4m^2c^2}\right].
\label{ata}
\eea
Eliminating $D_{\alpha\beta}(\q_1, \q_2)$ from (\ref{DA}) we obtain
linear homogeneous equation on the formfactor $F^{AA}_{\alpha\beta}$ with
degenerated kernel:
\bea
F^{AA}_{\alpha\beta}(\q_1, \q_2) &=&
\lambda \frac{2}{(2\pi)^3}\int d^3k_1d^3k_2
\delta (\k_1 + \k_2 - \p)\times \nonumber \\
&\times&\frac{F^{AA}_{\alpha\beta}(\k_1, \k_2)}
{E_A(\k_1) + E_A(\k_2) - \mu (\p)}
\left[1 + \frac{(\q_1 + \k_1)(\q_2 + \k_2)}{4m^2c^2}\right].
\label{aa}
\eea
Analogous equation can be derived for the formfactor $F^{A\tilde{A}}$
as well.

Let us pass to variables  $\k = \frac{1}{2}(\k_1 - \k_2),\;\;
\vec{Q} = (\k_1 + \k_2)$ in the integral and consider the equation
(\ref{aa}) at $\p = 0$ i.e., the bound state rest-frame case.
Then
\be
F^{AA}_{\alpha\beta}(\k) =
\lambda \frac{2}{(2\pi)^3}\int d^3q
\frac{F^{AA}_{\alpha\beta}(\q)}
{2E_A(\q)  - \mu_A(0)}
\left(1 + \frac{(\k^2 + \q^2)}{4m^2c^2}
 + \frac{(\q\cdot\k)}{4m^2c^2}\right),
\label{F0}
\ee
where $\mu_A(0)$ is an energy "gap"  in the bound state $\mid AA>$
spectrum.
\be
F^{AA}_{\alpha\beta}(\k) = A_{\alpha\beta} + \k^2B_{\alpha\beta} +
\k \vec{C}_{\alpha\beta}.  \label{eq1}
\ee
Here $A_{\alpha\beta}, B_{\alpha\beta}, \vec{C}_{\alpha\beta}$ are constant
matrices. Skewsymmetric $A_{\alpha\beta}$ and $B_{\alpha\beta}$ and
symmetric $\vec{C}_{\alpha\beta}$ over $\alpha,\; \beta$ matrices contribute
independently to the bound states and correspond to isoscalar and isovector
states.
Therefore, $A_{\alpha\beta} = A\epsilon_{\alpha\beta}, \;\;\;  B_{\alpha\beta}
=
B\epsilon_{\alpha\beta}$ and $\vec{C}_{\alpha\beta}$ can be expanded
over three symmetric matices: $I,\;\tau_1,\;\;\tau_3$. According to
these remarks the equations (\ref{F0})  are brought to the following set
of equations:
\bea
A &=& \lambda \frac{2}{(2\pi)^3}\int d^3k\left(1 + \frac{\k^2}
{4m^2c^2} \right)\frac{A + \k^2B}{2E(\k) - \mu_s}\nonumber \\
B &=& \lambda \frac{2}{(2\pi)^3}\int d^3k \frac{1}
{4m^2c^2} \frac{A + \k^2B}{2E(\k) - \mu_s}\nonumber \\
C^i_{\alpha\beta} &=& \lambda \frac{2}{(2\pi)^3}\int d^3k
\frac{2k^ik^j}{4m^2c^2} \frac{C^j_{\alpha\beta}}{2E(\k) - \mu_v}. \label{eq2}
\eea

Here $\mu_s, \;\mu_v \equiv \mu_A(0)$ stand for isoscalar and isovector states
accordingly. We have suppressed earlier the index $"A"$  on energy spectra
and $\mu (0)$, because these relations are fair for the state
$\mid \tilde{A} \tilde{A}>$ with the respective replacement of the energy
specta; and for the state $\mid A \tilde{A}>$ according to the remark after
(\ref{bs})), sign in front of 1 in the integral has to be changed.

 From the last relation we obtain usual equation to detrmine the "gap"
$\mu_v$ of isovector state:
\be
1 = \frac{4\lambda}{3}\frac{1}{(2\pi)^3}\int d^3k \frac{\k^2}{4m^2c^2}
    \frac{1}{2E(\k) - \mu_v}\label{muv}
\ee
First two equations form linear homogeneous system in respect to
$A$ and $B$. Demanding the determinant of this system to be zero
we come to the equation on
$\mu_s$:
\be
\left(\mbox{I}_1 - 1\right)\left(\mbox{I}_4 - 1\right) - \mbox{I}_2
\mbox{I}_3 = 0, \label{ge}
\ee
where
\bea
\mbox{I}_1
 &=& \lambda \frac{2}{(2\pi)^3}\int d^3k\left(1 + \frac{\k^2}
{4m^2c^2} \right)\frac{1}{2E(\k) - \mu_s}\nonumber \\
\mbox{I}_2
 &=& \lambda \frac{2}{(2\pi)^3}\int d^3k\left(1 + \frac{\k^2}
{4m^2c^2} \right)\frac{\k^2}{2E(\k) - \mu_s}\label{I} \\
\mbox{I}_3
 &=& \lambda \frac{2}{(2\pi)^3}\int d^3k \frac{1}
{4m^2c^2} \frac{1}{2E(\k) - \mu_s}\nonumber \\
\mbox{I}_4
 &=& \lambda \frac{2}{(2\pi)^3}\int d^3k \frac{\k^2}
{4m^2c^2} \frac{1}{2E(\k) - \mu_s},\nonumber
\eea
$\mu_s$ is an energy "gap" of isoscalar state.
We will make analysis of equation (\ref{ge}) and energy spectra
$E_A(\k)$ and $E_{\tilde{A}}(\k)$ (\ref{E2}) after the nature of splitting
of energy
"gaps" and masses of $A$ and $\tilde{A}$  particles will be studied.
Here we just note that if one inputs parameter
$\chi^2$:
\be
\chi^2 = M(2E(0) - \mu_s), \label{Ka}
\ee
then from (\ref{ge}) the equation on it follows:
\bea
&& \lambda \frac{M}{(2\pi)^3}\int \limits^\Lambda \frac{d^3k}{\k^2 +
\chi^2} = \label{eq3} \\
=&& \left(\frac{M g}{4m^2c^2} - \frac{1}{2}\right)^2\left[\frac{1}{2} -
\frac{\chi^2}{4m^2c^2} + \frac{M g}{4m^2c^2}\cdot
\frac{<\k^2> + \chi^2}{4m^2c^2}\right]^{-1}
\nonumber
\eea
\vspace{.5cm}
\centerline{ \bf 5. Symmetries of the model.}

\vspace{.5cm}

The model at hand apart from the trivial isotopic and $U(1)$ symmetries
has one more invariance, for the Hamiltonian (\ref{H}) on the solutions
$f^a(\k) = const, g^a(\k) = const$ does not depend on these amplitudes,
though the fields $\Psi^a_\alpha (\x, 0)$ depend on them. Thus, according
to Nether theorem, there should be conserved currents generated by variations
$\delta f^a$ and $\delta g^a$. The amplitudes $f^a(\k)$ and $g^a(\k)$ on
equations
(\ref{con}) are determined by three independent parameters, and this
parametrization can be chosen by many ways. The most simple is
\be
f^{a} = e^{i\varphi}
     \left( \begin{array}{c}
             e^{i\psi}\cos\theta \\
            -e^{-i\psi}\sin\theta
              \end{array}  \right) ,\;\;
g^{a} = e^{-i\varphi}
     \left( \begin{array}{c}
             e^{i\psi}\sin\theta \\
             - e^{-i\psi}\cos\theta
              \end{array}  \right).
\label{c}
\ee
Varying these relations over $\psi, \varphi$ and $\theta$ we obtain
variations of the fields $\Psi^a_\alpha (\x, 0)$:
\bea
\delta_\theta\Psi^a_\alpha (\x) &=& \frac{1}{(2\pi)^\frac{3}{2}}\int d^3k
\left[-g^a e^{2i\varphi}e^{i\k\x}A_\alpha(\k) + f^a e^{-2i\varphi}
e^{-i\k\x}\tilde{A}^\dagger_\alpha(\k)\right]\delta\theta \nonumber \\
\delta_\psi\Psi^a_\alpha (\x) &=& \frac{1}{(2\pi)^\frac{3}{2}}\int d^3k
\left[-i\sigma_{ab}\bar{g}^b e^{i\k\x}A_\alpha(\k) + i\sigma_{ab}\bar{f}^b
e^{-i\k\x}\tilde{A}^\dagger_\alpha(\k)\right]\delta\psi  \nonumber \\
\delta_\varphi\Psi^a_\alpha (\x) &=& \frac{1}{(2\pi)^\frac{3}{2}}\int d^3k
\left[if^a e^{i\k\x}A_\alpha(\k) - ig^a
e^{-i\k\x}\tilde{A}^\dagger_\alpha(\k)\right]\delta\varphi, \label{var}
\eea
where $\sigma_{ab} = \left(\begin{array}{cc} 0 & 1 \\
1 & 0 \end{array}\right)_{ab}$.\vspace{0.4cm}
Further on, by conventional method we find corresponding charges:
\bea
\hat{Q}_1 &=& \frac{i}{2}\int d^3k\left[e^{i\omega} A^{\dagger
}_\alpha (\k) \tilde{A}^{\dagger}_\alpha (-\k)
 - e^{-i\omega} \tilde{A}_\alpha
(-\k) A_\alpha (\k)\right]\nonumber \\
\hat{Q}_2 &=& \frac{1}{2}\int d^3k\left[e^{i\omega}
A^{\dagger}_\alpha (\k) \tilde{A}^{\dagger}_\alpha (-\k) + e^{-i\omega}
\tilde{A}_\alpha (-\k)A_\alpha (\k)\right]\nonumber \\
\hat{Q}_3 &=&
\frac{1}{2}\int d^3k\left[A^{\dagger
}_\alpha (\k) A_\alpha (\k) - \tilde{A}_\alpha (\k)
\tilde{A}^{\dagger}_\alpha (\k)\right] \label{gen}
\eea
where $\omega$ is an arbitrary phase. Here and in what follows
$\Psi_\alpha^a(\x)\equiv \Psi_\alpha^a(\x, 0)$.

Direct calculations result in the following commutation relations
\be
\left[\hat{Q}_i, \hat{Q}_j\right] = i \epsilon_{ijk}\hat{Q}_k,\;\;\;
\left[H, \hat{Q}_i\right] = 0,\;\;\;\left[\hat{Q}_i, \hat{T}_j\right] = 0,\;\;
\;\left[\hat{T}_i, \hat{T}_j\right] = i \epsilon_{ijk}\hat{T}_k ;
\label{alg}
\ee
$\hat{T}_i$ are generators of isotopic transformations, defined as
\bea
\hat{T}_i &=& \frac{1}{2}\int d^3x \Psi^{\dagger a}_\alpha (\x)\tau^i_{\alpha
\beta}\Psi^a_\beta (\x) = \nonumber \\
&=& \frac{1}{2}\int d^3k \tau^i_{\alpha \beta}\left[A^{\dagger
}_\alpha (\k) A_\beta (\k) - \tilde{A}_\alpha (\k)
\tilde{A}^{\dagger}_\beta (\k)\right]
\label{iso}
\eea
To the above charges, the $U(1)$ charge is to be added:
\be
\hat{Q}_{U(1)} = \int d^3k\left(A^{\dagger
}_\alpha (\k) A_\alpha (\k) + \tilde{A}_\alpha (\k)
\tilde{A}^{\dagger}_\alpha (\k)\right)
\label{phase}
\ee
Thus, we have seven charges that exhaust the whole symmetry of the model.
This symmetry forms $su(2)_T\times su(2)_Q$ and $u(1)$ algebra
(\ref{alg}).

Let us show now that group of unitary transformations with generators
$\hat{Q}_i$ leaves the form of the Hamiltonian (\ref{H}) invariant.
Defining group element $U(\alpha ,\beta ,\gamma )$
\be
U(\alpha ,\beta ,\gamma ) = e^{i\gamma\hat{Q}_3}e^{i\beta\hat{Q}_2}
e^{i\alpha\hat{Q}_1}, \label{gr}
\ee
consider thransformation of the field $\Psi_\alpha^a(\x)$, calculating
first transformations of operators $A_\alpha (\k)$ and
$\tilde{A}^\dagger_\alpha (\k)$. Let
\bea
a_\alpha (\k) &=& U^\dagger (\alpha ,\beta ,\gamma )A_\alpha (\k)
U(\alpha ,\beta ,\gamma )\nonumber \\
\tilde{a}_\alpha (\k) &=& U^\dagger (\alpha ,\beta ,\gamma )
\tilde{A}_\alpha (\k) U(\alpha ,\beta ,\gamma ). \label{trans}
\eea
After the simple calculations we have:
\bea
a_\alpha (\k) &=& e^{i\gamma}\left(\cos\alpha  \cos\beta - i\sin\alpha
\sin\beta\right)A_\alpha (\k) - \nonumber \\
&-&e^{i(\omega - \gamma)}\left(\sin\alpha
\cos\beta - i\cos\alpha \sin\beta\right) \tilde{A}^\dagger_\alpha (-\k)
\nonumber \\
\tilde{a}_\alpha (-\k) &=& e^{-i\gamma}\left(\cos\alpha  \cos\beta +
i\sin\alpha \sin\beta\right)\tilde{A}^\dagger_\alpha (-\k) + \nonumber \\
&+&e^{-i(\omega - \gamma)}\left(\sin\alpha\cos\beta +
i\cos\alpha \sin\beta\right) A_\alpha (\k) \label{trans1}
\eea
 From what follows reverse transformations
\bea
A_\alpha (\k) &=& e^{-i\gamma}\left(\cos\alpha  \cos\beta + i\sin\alpha
\sin\beta\right)a_\alpha (\k) + \nonumber \\
&+&e^{i(\omega - \gamma)}\left(\sin\alpha
\cos\beta - i\cos\alpha \sin\beta\right) \tilde{a}^\dagger_\alpha (-\k)
\nonumber \\
\tilde{A}^\dagger_\alpha (-\k)
 &=& e^{i\gamma}\left(\cos\alpha  \cos\beta -
i\sin\alpha \sin\beta\right)\tilde{a}^\dagger_\alpha (-\k) -\nonumber \\
&-&e^{-i(\omega - \gamma)}\left(\sin\alpha\cos\beta +
i\cos\alpha \sin\beta\right) a_\alpha (\k) \label{trans2}
\eea
Now, using (\ref{trans1}), we obtain:
\bea
\Psi_\alpha^a(\x) &=& \frac{i}{2}\int d^3k\left[ f^a e^{i\k\x}A_\alpha (\k)
+ g^a e^{-i\k\x}\tilde{A}^\dagger_\alpha (\k)\right] = \nonumber \\
 &=& \frac{i}{2}\int d^3k\left[ N^a e^{i\k\x}a_\alpha (\k)
+ M^a e^{-i\k\x}\tilde{a}^\dagger_\alpha (\k)\right], \mbox{where}
\label{transf} \\
N^a &=& e^{-i\gamma}\left(\cos\alpha  \cos\beta + i\sin\alpha
\sin\beta\right)f^a - \nonumber \\
&-& e^{-i(\omega - \gamma)}\left(\sin\alpha
\cos\beta + i\cos\alpha \sin\beta\right)g^a \nonumber \\
M^a &=& e^{i\gamma}\left(\cos\alpha  \cos\beta -
i\sin\alpha \sin\beta\right) g^a + \nonumber \\
&+& e^{i(\omega - \gamma)}\left(\sin\alpha\cos\beta -
i\cos\alpha \sin\beta\right)f^a .\nonumber
\eea
By the strightforward calculation one can check that
\be
\sum \limits_a N^a\bar{N}^a = \sum \limits_a M^a\bar{M}^a = 1 ,\;\;\;
\sum \limits_a N^a\bar{M}^a = 0  \label{equiv}
\ee
Thus, the amplitudes $N^a$ and $M^a$ have the same properties as the
amplitudes $f^a$ and $g^a$. Therefore, the Hamiltonian written in terms
of $a_\alpha (\k)$ $\tilde{a}_\alpha (\k)$ will have the same form i.e.,
(\ref{trans1}) and (\ref{trans2}) do not change the form
oh the Hamiltonian. Hence the  imortant conclusion follows: the point is
that the transformations (\ref{trans1}) and (\ref{trans2}) are unitary -
inequivalent as it follows from this same form of the generators
$\hat{Q}_1$ and $\hat{Q}_2$. Consequently, the parameters $\alpha ,\beta ,
\gamma$ fix different Hilbert spaces orthogonal to each other.
The forminvariance of the Hamiltonian means that, although the pointed
transformations are unitary - inequivalent, the dynamical reconstruction
does not take place i.e., vacuum energy density, $n$ - particle spectra
and {\sl ect}, are the same in all Hilbert spaces.

The next impotant aspect of $SU(2)_Q$ - invariance is that in Hilbert space
constructed by means of operators $A^\dagger_\alpha (\k)$ and
$\tilde{A}^\dagger_\alpha (\k)$, this symmetry occurs to be spontaneously
broken. Indeed, for the vacuum expectations of generators
$\hat{Q}_i$ we have:
\be
\left< 0 \mid \right.\hat{Q}_1 \vac_{A\tilde{A}} =
\left< 0 \mid \right.\hat{Q}_2 \vac_{A\tilde{A}} = 0\;\;\;
\left< 0 \mid \right.\hat{Q}_3 \vac_{A\tilde{A}} = - \frac{V}{V^*}
\label{break}
\ee
where $V$ is a space volume, and $V^*$ is defined by (\ref{def}).
The last relation indicates that $SU(2)_Q$ symmetry is spontaneously
broken. Condensate $\hat{Q}_3$ is microscopic object i.e., it is
proportional to the space volume. That, in its turn, is connected
with the state $\hat{Q}_3 \vac_{A\tilde{A}}$ to be normless \cite{grib1}.

Let us consider now, how to classify the states in regard of
$SU(2)_Q$. The spectrum of $SU(2)_Q$ Casismir operator is know to be
\bea
&&Q_1^2 + Q_2^2 + Q_3^2 = L(L + 1), \;\;L = 1, \frac{1}{2}, 1
\dots ; \nonumber \\
&&Q_3 = -L, -L + 1,\dots L, \label{kasimir}
\eea
increasing and decreasing operators $\hat{Q}_\pm = \hat{Q}_1 \pm i
\hat{Q}_2$ is written as
\bea
\hat{Q}_+ &=& i\int d^3k e^{i\omega} A^{\dagger
}_\alpha (\k) \tilde{A}^{\dagger}_\alpha (-\k) \nonumber \\
\hat{Q}_- &=& -i\int d^3k e^{-i\omega} \tilde{A}_\alpha (-\k)A_\alpha (\k)
\label{incr}
\eea
As is seen from (\ref{incr}), action of
$\hat{Q}_+$ to the vacuum  increases the number of pairs $A\tilde{A}$. Further
we
have:
\bea
&&\hat{Q}^2\vac = \frac{V}{V^*}\left(\frac{V}{V^*} + 1\right)\vac,\;\;
\hat{Q}^2\left(\hat{Q}_+\right)^n\vac = \frac{V}{V^*}\left(\frac{V}{V^*} +
1\right)
\left(\hat{Q}_+\right)^n\vac, \;\;\mbox{with} \nonumber \\
&&\hat{Q}_3
\left(\hat{Q}_+\right)^n\vac = \left(-\frac{V}{V^*} + n\right)
\left(\hat{Q}_+\right)^n\vac,\;\;\hat{Q}_-\vac = 0.
\label{multip}
\eea
Thus the vacuum  and all states with n number of excited pairs
$A\tilde{A}$ lie in one the same $SU(2)_Q$ multiplet with
$Q^2 = \frac{V}{V^*}\left(\frac{V}{V^*} + 1\right)$ and dimension
$N = 2\frac{V}{V^*} + 1$. Let us call it "vacuum multiplet".

 From (\ref{kasimir}) follows $\frac{V}{V^*} =
1/2, 1, 3/2, \dots$. Addition of pair $A\tilde{A}$ increases a maximum value
of the projection $Q_3$ and its maximal value is acheived at
$n = 2\frac{V}{V^*}$. Next, since $V \sim \infty$, the vacuum and all
the pairs $A\tilde{A}$ lie in the right, infinite end of spectrum
(\ref{kasimir}). What states correspond to finite - dimensional
representations of $SU(2)_Q$? In order to clarify this question
consider one - particle state $A^\dagger_\alpha (\k)\vac$
(analogous for $\tilde{A}^\dagger_\alpha (\k)\vac$).
Owing to the relation
$$
\hat{Q}_-A^\dagger_\alpha (\k)\vac = 0
$$
this state has minimal value of  $Q_3$
\be
\hat{Q}_3A^\dagger_\alpha (\k)\vac = \left(-\frac{V}{V^*} + \frac{1}{2}
\right)A^\dagger_\alpha (\k)\vac, \;\;\mbox{with}\;\;
Q^2 = \left(\frac{V}{V^*} - \frac{1}{2}\right)
      \left(\frac{V}{V^*} - \frac{1}{2} + 1 \right).
\label{onepm}
\ee
As follows from here the one-particle state lies in representation with
dimension $N = 2\frac{V}{V^*}$, whereas the dimension of vacuum
representation is equal to $N = 2\frac{V}{V^*} + 1$. It is easy to show
that the increasing of number of one kind of excitations ($A$ or $\tilde{A}$)
will lead to the consecutive decreasing of dimension of representation.
This suggests the way to construct a state in which $Q^2$ and $Q_3$ will
have finite values. The number of excitations, however, must be infinitely
large to cancel the infinite value $\frac{V}{V^*}$.
Evidently that the only chance to realize this program is a phase
transition accompanied by dynamical reconstraction of vacuum.
We will show in a moment that thus reconstructed vacuum and its excitations
realize finite-dimensional representations of $SU(2)_Q$ and, what is
more important, $SU(2)_Q$ symmetry becomes exact (restored).
There arises interesting picture: infinite - dimensional (vacuum)
representation is realized in the system with sontaneously broken symmetry,
whereas finite - dimensional $SU(2)_Q$ representation is realized in
the system with exact (restored) symmetry.
As a conclusion of this section note that excitations $A$ and $\tilde{A}$
have equal by value but different by sign $Q_{U(1)}$ charges and equal $Q_3$
charges.
\vspace{.5cm}

\centerline{\bf 6.Realization of finite - dimensional $SU(2)_Q$
representations.}
\vspace{.5cm}

The above set problem can be solved by introduction of such a Bogolubov
transformations that would have `saturated' vacuum  by particles of one
kind. As a sample of these paticles we take $\tilde{A}$, define
hermitian generator:
$\hat{Q}_{\tilde{A}}$:
\be
\hat{Q}_{\tilde{A}} = \frac{i}{2}\int d^3k\epsilon_{\alpha\beta}
\left[ e^{i\phi}\tilde{A}^\dagger_\alpha (\k)
\tilde{A}^\dagger_\beta (-\k) + e^{-i\phi}
\tilde{A}_\alpha (\k)\tilde{A}_\alpha (-\k)\right]
\label{bogol}
\ee
and consider the transformations
\bea
\tilde{B}_\alpha (\k) &=& U^\dagger (\omega)\tilde{A}_\alpha (\k) U(\omega),
\;\;\;U(\omega) = e^{i\omega\hat{Q}_{\tilde{A}}} \nonumber \\
B_\alpha (\k) &=& A_\alpha (\k). \label{bogol1}
\eea
Define now the vacuum
$B_\alpha (\k)\Vac = \B_\alpha (\k)\Vac = 0$. All combination of
the transformations can be represented by the following scheme:
\be
A \rightarrow B,\;\;\; \A \rightarrow \B,\;\;\;\vac_{A\tilde{A}}\rightarrow
\Vac .\label{schem}
\ee
 From (\ref{bogol1}) we obtain:
\bea
\B_\alpha (\k) &=& \cos\omega \A_\alpha (\k) - e^{i\phi}\sin\omega \epsilon_{
\alpha \beta}\A^\dagger_\beta (-\k) \nonumber \\
B_\alpha (\k) &=& A_\alpha (\k) \label{bogol2} \\
\A_\alpha (\k) &=& \cos\omega \B_\alpha (\k) + e^{i\phi}\sin\omega \epsilon_{
\alpha \beta}\B_\beta (-\k) \nonumber \\
A_\alpha (\k) &=& B_\alpha (\k)
\eea
Note, that insertion into the generator $\hat{Q}_{\tilde{A}}$ (\ref{bogol})
skewsymmetric tensor $\epsilon_{\alpha \beta}$ leads to the condensing
in the vacuum of the pairs $\A\A$ in a state with zero isotopic spin.
If now one inputs into the Hamiltonian (\ref{two})  $A$ and $\tilde{A}$,
expressed via $B$ and $\B$ from (\ref{bogol2}) , then again the Hamiltonian
splits in to the normal and fluctuation parts. As was already argued,
the presnce of the latter destroys stability of vacuum and its excitations
i.e., they would not be the eigenstates of the Hamiltonian.
Nontrivial rotation on angle $\omega = \pi /2$ is relevant to the case
when fluctuation part is absent and the Hamiltonian in terms of $B$ and $\B$
has the form
\bea
H &=& \int d^3k \left[E_B(\k)B^\dagger_\alpha (\k)B_\alpha (\k) +
E_{\B}(\k)\B^\dagger_\alpha (\k)\B_\alpha (\k)\right] + \nonumber \\
&+& \lambda \frac{1}{(2\pi)^3}\int d^3k_1d^3k_2d^3k_3d^3k_4
          \biggl\{\biggr. \delta(\k_1 - \k_2 + \k_3 - \k_4)\times \nonumber \\
&\times& \left[1 -
\frac{(\k_1 + \k_2)(\k_3 + \k_4)}{4m^2c^2}\right]\times \nonumber \\
&\times&\left[B^\dagger_\alpha (\k_1)B^\dagger_\beta (\k_3)
B_\alpha (\k_2)B_\beta (\k_4) + \B^\dagger_\alpha (\k_1)\B^\dagger_
\beta (\k_3)
\B_\alpha (\k_2)\B_\beta (\k_4) + \right.\nonumber \\
&+& \left.\left.2B^\dagger_\alpha (\k_1)\B^\dagger_
\beta (\k_3)B_\alpha (\k_2)\B_\beta (\k_4)\right]\right\},
\label{HB}
\eea
where
\be
E_B (\k) = E_{\B} (\k) = \varepsilon(\k) + g\dot\frac{k^2}{4m^2c^2} - g +
g\dot\frac{<k^2>}{4m^2c^2} \label{spectr}
\ee
For the charges $\hat{Q}_i$ and $\hat{Q}_{U(1)}$ after input of (\ref{bogol2})
into (\ref{gen}) follows:
\bea
\hat{Q}_1 &=& \frac{i}{2}\int d^3k\epsilon_{\alpha\beta}
\left[e^{i(\omega - \phi)} B^{\dagger
}_\alpha (\k) \tilde{B}_\beta (\k)
 - e^{-i(\omega - \phi)} \tilde{B}^\dagger_\beta
(\k) B_\alpha (\k)\right]\nonumber \\
\hat{Q}_2 &=& \frac{1}{2}\int d^3k\epsilon_{\alpha\beta}
\left[e^{i(\omega - \phi)}
B^{\dagger}_\alpha (\k) \tilde{B}_\beta (\k) + e^{-i(\omega - \phi)}
\tilde{B}^\dagger_\beta (\k)B_\alpha (\k)\right]\nonumber \\
\hat{Q}_3 &=&
\frac{1}{2}\int d^3k\left[B^{\dagger
}_\alpha (\k) B_\alpha (\k) - \tilde{B}^\dagger_\alpha (\k)
\tilde{B}_\alpha (\k)\right] \label{gen1} \\
\hat{Q}_{U(1)} &=& \int d^3k\left(B^{\dagger
}_\alpha (\k) B_\alpha (\k) + \tilde{B}^\dagger_\alpha (\k)
\tilde{B}_\alpha (\k)\right) \nonumber
\eea
 From expressions for the Hamiltonian (\ref{HB}) and charges
$\hat{Q}_i$ and $\hat{Q}_{U(1)}$ follows the relations pointing to the fact
that the transformations (\ref{bogol2}) really restore $SU(2)_{Q}$ symmetry,
namely:
\be
H\Vac = \hat{Q}_1\Vac  = \hat{Q}_2\Vac =\hat{Q}_3\Vac = \hat{Q}_{U(1)}\Vac =
0  \label{symm}
\ee
Here one is to add the degeneration of $B$ and $\B$ spectra (\ref{spectr}).
As can be seen from the condition $\hat{Q}_3\Vac = 0$
vacuum $\Vac$ lies in a singlet $SU(2)_{Q}$ representation,
and from
\be
\hat{Q}_3B^\dagger_\alpha (\k)\Vac  = +\frac{1}{2}B^\dagger_\alpha (\k)\Vac,
\;\;\;\hat{Q}_3\B^\dagger_\alpha (\k)\Vac  = -\frac{1}{2}
\B^\dagger_\alpha (\k)\Vac \label{Q3}
\ee
follows that one-particle excitations $B$ and $\B$ form fundamental
$SU(2)_{Q}$ representation.

The addition of the same sort of particles increases the dimension
of representation. Thus we obtain the result: states of the system with
unbroken symmetry realize finite - dimensional representations of $SU(2)_{Q}$.
At the
spontaneous symmetry breaking (scheme (\ref{schem})
in backword direction)
the vacuum and its excitations will realize the
representation from right infinite - dimensional end of the spectrum.

As is well known \cite{grib1}, the inequality to zero of vacuum expectation
value of some
generator commuting with Hamiltonian is not a sufficient sign of
spontaneous symmetry breaking. It is also necessary to have  a parameter
that regulates breaking and restoration of the symmetry. Critical value of
the parameter separates these two deifferent regions. Physical meaning of the
parameter can vary for different systems and processes. For example, the
temperature is the parameter in  superconductive theory,
the mass - in the scalar model $\phi^4$. In order to reveal the parameter
in our case we should answer the question: at what conditions it is
energetically preferable for the system in the state with unbroken symmetry
to pass to the state with spontaneously broken symmetry? and vice versa,
if the initial state of the system is the state with spontaneously broken
symmetry, when it is
energetically preferable to restore the symmetry? It is clear that an answer
follows from the investigation of the vacuum energy density. In the state
with spontaneously broken symmetry it is defined by the relation (\ref{W}),
and in the state with unbroken symmetry it is equal to zero (\ref{symm}).
The sign of the relation $<\varepsilon (\k)> - 2g$ is crucial for the answer.
However, to define the sign it is necessary to know the value
$<k^2>$. We will show that it can be calculated by implying certain
requirements on the one-particle spectrum (\ref{spectr}).
Let $E(\k)\equiv E_B(\k) = E_{\B}(\k)$ and has the following form:
\bea
E(\k) &=& \frac{k^2}{2M} + Mc^2 + \triangle E,\;\;\;
\triangle E = - Mc^2 + mc^2 - g + g\frac{<k^2>}{4m^2c^2}, \nonumber \\
M &=& \frac{m}{1 + \frac{g}{2mc^2}},\;\;\;
m = \frac{M}{2}\left[1 + \sqrt{1 + \frac{2g}{Mc^2}}\right] \label{Mm}
\eea

Hence it follows, that "physical" mass $B$ and $\B$ particles
coincides with mass
$m_A$ which is defined in (\ref{E3}). Now let us make a statement henceforth
important. Since the vacuum $\Vac$ has all the quantum numbers equal
to zero we will require for the spectrum $E(\k)$ to describe the "normal"
nonrelativistic particle with mass $M$
i.e., we will require $\triangle E$ to be equal to zero, for there are no
physical reasons for its existence. This condition determines $<k^2>$ via
the renormalized mass $M$ and via the coupling constant $g$:
\bea
\frac{<k^2>}{4m^2c^2} &=& 1 + \frac{Mc^2}{2g}\left[1 - \sqrt{1
+ \frac{2g}{Mc^2}} \right] = \frac{\sqrt{1 + G}}{1 + \sqrt{1 + G}},
\nonumber \\
\frac{<k^2>}{2m} &=& Mc^2\sqrt{1 + G} \Rightarrow  <k^2> = M^2c^2 \left[1 + G +
\sqrt{1 + G}\right], \;\;\;\mbox{with} \label{triang0} \\
G &=& \frac{2g}{Mc^2},\;\;\;\triangle E = 0 \nonumber
\eea
The value of $<k^2>$ characterizes the fluctuation of the momentum inside
the excitation. The radius $R$ of the localization area $R$ is connected
with it by the uncertainty  relation  $R^2<k^2>\simeq 1$.
 From the expression for
$<k^2>$ it is clear that at infinitesimal $G$ the radius $R$ is defined by
compton length of the excitation i.e., $R \sim \hbar /Mc$; which is rather
reasonable result.

The cut-off parameter $\Lambda$ is expressed from the relation (\ref{def})
via renormalized mass $M$ and constant $G$. It is interesting to note that
although the cut-off is not invariant procedure, the final expression for
$\Lambda$, obtained from condition $\triangle E = 0$, includes only invariant
quantities. It is also interesting to write down the expressions for
$\Lambda$ and renormalized coupling constant $g$ as functions of "bare" mass
$m$ and constant $\lambda$:
\bea
g &=& 2mc^2\alpha_0\left[1 + 9\alpha_0 + \frac{195}{2}\alpha_0^2 + \cdots
\right], \nonumber \\
\Lambda &=& \sqrt{\frac{10}{3}}mc\left[1 + 3\alpha_0 + \frac{47}{2}
\alpha_0^2 + \cdots\right]; \label{Lg} \\
\alpha_0 &=& \left(\frac{10}{3}\right)^{\frac{3}{2}}\cdot\frac{\lambda m^2c}
{12\pi^2}. \nonumber
\eea
It is clear that the coefficients of the expansion encrease assymtotically,
what is probably reflects the nonrenormalizability of the model.

Now when $<k^2>$ is known we are ready to express the vacuum energy density
(\ref{W}) via "physical" mass $M$ and constant $G$. So we have:
\be
\frac{V}{V^*}W_0 = Mc^2\left[3\sqrt{1 + G} + 1 - 2G\right]. \label{W1}
\ee
Recall that this is energy density of vacuum $\vac_{A\A}$ i.e., of the
system with broken $SU(2)_{Q}$ symmetry. If the r.h.s. of (\ref{W1}) is
positive, then it is energetically preferable for the system to pass
spontaneously to the state $\Vac$ where energy is equal to zero.
If the r.h.s. of (\ref{W1}) is negative then for the system  in a state
with unbroken symmetry it is energetically preferable to pass to the
state with spontaneously broken symmetry. And the dimensionless coupling
constant $G$ is the mensioned parameter that regulates regime of spontaneous
transitions in our model. The critical value $G_{cr}$ is the one at which
the vacuum energy density (\ref{W1}) vanishes.
\be
G_{cr}^2 -\frac{13}{4}G_{cr} - 2 = 0 \label{gcr}
\ee
 From here we find: $G_{cr} = \frac{2g^{cr}}{Mc^2} \simeq 3.75$.
Recall that the "piercing" of vacuum takes place at $G = 8$ i.e., really
in the region where the symmetry is spontaneously broken.

Now, let us return to the spectra $E_A(\k)$ and $E_{\A}(\k)$ (relations
(\ref{E2}), (\ref{E3}), (\ref{E4}), (\ref{EE})). First, comparing
(\ref{spectr}) and (\ref{E2}) we obtain: $m_A = M$. Further, for energy
"gaps" we find the following expression:
\bea
E_A(0) &=& Mc^2 - 4g = Mc^2(1 - 2G), \nonumber \\
E_{\A}(0) &=& = Mc^2(2G - \sqrt{1 + G}) \label{split}
\eea
It is worth mensioning that for the "normal" spectrum i.e., for
the excitation spectrum of vacuum without condensate, the energy "gap"
should coincide with the exitation mass, as it happens for
$B$ and $\B$ particles. However, at the presence of condensates, always
arising at spontaneous symmetry breaking it is not true, what can be seen
from example of the spectrum $E_A(\k)$ in (\ref{split}). To the
natural width of the "gap" $Mc^2$ appends the term generated by  spontaneous
transition. Thus, besides the renormalization of mass, related with "dressing"
of particle because of the interaction, there also takes place
renormalization of energy "gap", caused by spontaneous transition i.e.,
by dynamical reconstruction of vacuum. The energy "gap" $E_{\A}(0)$
is difficult generally to interpret, for its the interpretation as a real
particle is closely connected with the "piercing" of vacuum.
\vspace{.5cm}

\centerline{\bf 7. Bound states of $B$ and $\B$ excitations}.
\vspace{.5cm}

As shows the consideration of bound states $A$ and $\A$ particles,
to obtain equation on wave function and mass of bound state one needs to
act by the Hamiltonian to the two particle state. In terms of $B$ and $\B$
excitations the Hamiltonian is expressed by (\ref{HB}). Its action to
the two - particle state yields:
\bea
&&H B^\dagger_\alpha (\q_1)B^\dagger_\beta (\q_2)\Vac =  \int d^3k_1d^3k_2
\biggl\{\delta(\k_1+\q_1)\delta(\k_2+\q_2)\left[E(\k_1) + E(\k_2)\right] -
\nonumber \\
&&- \lambda \frac{2}{(2\pi)^{\frac{3}{2}}}
\delta(\k_1+\k_2 -\q_1-\q_2)
\left.\left[1 -\frac{(\k_1 + \q_1)(\k_2 + \q_2)}{4m^2c^2}\right]
\right\}B^\dagger_\alpha (\k_1)B^\dagger_\beta (\k_2)\Vac ; \nonumber \\
&&H \B^\dagger_\alpha (\q_1)\B^\dagger_\beta (\q_2)\Vac =  \int d^3k_1d^3k_2
\biggl\{\delta(\k_1+\q_1)\delta(\k_2+\q_2)\left[E(\k_1) + E(\k_2)\right] -
\nonumber \\
&&- \lambda \frac{2}{(2\pi)^{\frac{3}{2}}}
\delta(\k_1+\k_2 -\q_1-\q_2)
\left.\left[1 -\frac{(\k_1 + \q_1)(\k_2 + \q_2)}{4m^2c^2}\right]
\right\}\B^\dagger_\alpha (\k_1)\B^\dagger_\beta (\k_2)\Vac ; \nonumber \\
&&H B^\dagger_\alpha (\q_1)\B^\dagger_\beta (\q_2)\Vac =  \int d^3k_1d^3k_2
\biggl\{\delta(\k_1+\q_1)\delta(\k_2+\q_2)\left[E(\k_1) + E(\k_2)\right] -
\nonumber \\
&&- \lambda \frac{2}{(2\pi)^{\frac{3}{2}}}
\delta(\k_1+\k_2 -\q_1-\q_2)
\left.\left[1 -\frac{(\k_1 + \q_1)(\k_2 + \q_2)}{4m^2c^2}\right]
\right\}B^\dagger_\alpha (\k_1)\B^\dagger_\beta (\k_2)\Vac ;\nonumber \\
&&E(\q) = \frac{q^2}{2M} + Mc^2 \label{bbs}
\eea
Thus, $BB$, $\B\B$, $B\B$ pairs interact by the same way, have the same wave
functions and the masses of bound states. This picture corresponds to the
exact (unbroken) symmetry of the model. Equations on the formfactor
$F_{\alpha\beta}(\q_1,\q_2)$ and the mass of bound state,
as is seen from last relations,  coincide (up to
the one - particle spectra) with the equations (\ref{DA}), (\ref{eq2}),
(\ref{muv}). So, we make use of these previous results and write down
at once the equations on
$\mu_s(0)$ and $\mu_v(0)$ - isoscalar and isovector masses respectively:
\be
1 = \lambda \frac{1}{(2\pi)^3}\frac{M}{3m^2c^2}
\int \limits^\Lambda d^3k \frac{\k^2}
    {k^2 + \chi_v^2}   \label{bmuv}
\ee
where $\chi_v^2 = M\bigl(2Mc^2 - \mu_v(0)\bigr)$. \\
This equation after replacement of variables can be brought to the following
form:
\be
1 = \frac{2G}{(1 + \sqrt{1 + G})^2}\int \limits_0^1 \frac{t^4dt}{t^2 +
\bigl(\frac{\chi_v}{\Lambda}\bigr)^2} \label{nogo}
\ee
At $\chi_v = 0,\;\;G = \infty$ the r.h.s of this equation acheives its
maximal value equal to $2/3$. Therefore, this equation has no solution
at any $G$ i.e., there is no the bound state of two fermions with
the same helicity in the isovector state on the sector with the unbroken
$SU(2)_{Q}$ symmetry for the Lagrangian (\ref{L}).

After intergation in the l.h.s. of equation (\ref{eq3}) and simple
transformations for isoscalar $\chi_s^2 = M\bigl(2Mc^2 - \mu_s(0)\bigr)$,
we derive transcendental equation:
\be
\bigl(z^2c_1 - c_2\bigr)\bigl(z - \arctan z\bigr) = z^3, \label{tan}
\ee
where
$$
c_1 = \frac{9}{20}\cdot\frac{3 + 2G + \sqrt{1 + G}}{1 + G + \sqrt{1 + G}},
\;\;\;c_2 = \frac{3}{4}G\biggl(1 + \frac{2}{1 + \sqrt{1 + G}}\biggr),\;\;\;
z = \frac{\Lambda}{\chi_s}
$$
The analysis of the transcendental equation shows that it always has a
solution at $c_1 > 1$. With a good accuracy  $c_1 \simeq \frac{9}{10}G$,
therefore, the condition for the existence of a solution reduces to demand
$G > \frac{10}{9}$, and isoscalar state really lies on the sector with
unbroken $SU(2)_{Q}$ symmetry. Table 1 contain the results of numerical
solution of the equation (\ref{tan}). From it follows that $z(G)$
strongly changes in the region $\frac{10}{9} < G \leq 1.3$, but
further on, at $G\geq 1.3$, is slowly achieving its asymptotic  value
$z(\infty) = \sqrt{\frac{5}{6}}$.
\vspace{.5cm}

\centerline{\bf 8. Discussion of the results.}
\vspace{.5cm}

We have considered quantum field model of "singlet" fermions with isotopic
spin equal to 1/2 and with contact current$\times$current interaction.
This model represents non-relativistic limit of chiral Lagrangian, in which
the quantum numbers of current correspond to $\omega$-meson. Besides the
isotopic the model possesses additional degrees of freedom, related to the
existence of two solutions, which are realized in different Hilbert spaces.
The attempt to describe these solutions by a single canonically quantized
field, leads to the two component theory and the index numbering these
components is carried not by creation and annihilation operators, but
by the amplitudes at them. Such model has additional (to isotopic and
$U(1)$) $SU(2)_Q$ symmetry, which could be spontaneously broken or
exact with respect to the value of the dimensionless coupling constant
$G = \frac{2g}{Mc^2}$. Critical value $G_{cr} \simeq 3.75$
divides these two realizations of $SU(2)_Q$, and one-particle excitations are
essentially different for each realization. Moreover, at the phase transition
the dimensions of $SU(2)_Q$ multiplets (corresponding vacuums and their
excitations) are changed by "leap". The dynamical "dressing"
of fermionic masses via interaction in spontaneously broken region is
accompanied by the renormalization of energy "gaps" of corresponding
spectra.
It is worth noting also that the "physical" mass of $A,B,\B$ particles
always less then "bare" (current) mass $m$, and when the latter is
equal to zero, the "physical" mass is vanishing too. This picture is
essentially different from the one obtained by Numbu and Jiona-Lasinio,
where at $m = 0$ the non - zero "physical" mass is still left.
The Hartry-Fock method, they have used, conceals the circumstance
that there takes place, in fact, the renormalization of the energy "gap".
The calculation of "dressing" via interaction was made later on (see e.g.
\cite{domitr}).

The  excitation $\A$ arising at the $SU(2)_Q$ spontaneous breaking,
possesses a set of exotic properties. In dependence of the constant
$G$ value it could describe ether "bubble" in the vacuum or real particle.
(at $G >8$) with a mass singularly depending on constant $G$.
At $G = 8 + \epsilon$ where $\epsilon >0$ is an infinitesimal quantity,
the mass $m_{\A}$ could be as large as possible, and at any
 $G >8$ the excitation $\A$ is always heavier then excitation $A$ and
only at the limit $G\rightarrow \infty$ their masses become equal.

The further problem we would like to consider is construction of
dynamical mapping of heisenberg fields $\Psi (\x,t)$ on
"physical" fields $\Psi (\x,t=0)
\equiv \Psi (\x)$, which excitations are the eigenstates of the total
Hamiltonian. The solution of this problem will be done in a following
paper.
\vspace{2cm}
\begin{center}
Table 1. \\
\begin{tabular}{c|c}   \hline
G      &    z(G)     \\ \hline
1.2    &    175      \\
1.3    &    18.5     \\
1.4    &    10.5     \\
2.0    &    3.8      \\
3.0    &    2.7      \\
4.0    &    2.0      \\
5.0    &    1.85     \\
6.0    &    1.7      \\
7.0    &    1.65     \\
8.0    &    1.55     \\
9.0    &    1.48     \\
10.0   &    1.44     \\
\end{tabular}
\end{center}

\end{document}